\documentclass[prl,twocolumn,amsmath,amssymb,showpacs]{revtex4}

\usepackage{graphicx}
\usepackage{dcolumn}
\usepackage{bm}

\begin{document}

\preprint{APS/V. Sih}

\title{Generating Spin Currents in Semiconductors with the Spin Hall Effect}

\author{V. Sih}
\author{W. H. Lau}
\author{R. C. Myers}
\author{V. R. Horowitz}
\author{A. C. Gossard}
\author{D. D. Awschalom}
\email{awsch@physics.ucsb.edu}
\affiliation{%
Center for Spintronics and Quantum Computation\\
University of California, Santa Barbara, CA 93106}

\date{\today}

\begin{abstract}
We investigate electrically-induced spin currents generated by the
spin Hall effect in GaAs structures that distinguish edge effects
from spin transport. Using Kerr rotation microscopy to image the
spin polarization, we demonstrate that the observed spin
accumulation is due to a transverse bulk electron spin current,
which can drive spin polarization nearly 40 microns into a region
in which there is minimal electric field. Using a model that
incorporates the effects of spin drift, we determine the
transverse spin drift velocity from the magnetic field dependence
of the spin polarization.

\end{abstract}


\maketitle

The spin Hall effect refers to the generation of a spin current
transverse to a charge current in non-magnetic systems in the
absence of an applied magnetic field.  This spin current may arise
from spin-dependent scattering~\cite{dyakonov,hirsch} or from the
effect of the spin-orbit interaction on the band
structure~\cite{murakami,sinova}. Although spin current is
difficult to measure directly, the spin Hall effect was predicted
to create spin accumulation at the edges of a
channel~\cite{zhang}. This electrically-induced spin polarization
was observed in bulk epilayers of electron-doped
semiconductors~\cite{katoSHE} and in two-dimensional
hole~\cite{wunderlich} and electron systems~\cite{sih}, and recent
calculations~\cite{engel,tse06,nomura} show reasonable agreement
with the experimental results. However, determining the spin
current through analysis of the spin accumulation is complicated
because spin is not a conserved quantity in the presence of the
spin-orbit interaction~\cite{rashba,shi} and the choice of
boundary conditions has a strong effect on the calculated spin
accumulation~\cite{tse05,galitski}. In addition, it is possible
that spin polarization observed at the sample edges could be due
to an edge effect, such as depletion near the edge resulting in a
local spin splitting from the Bychkov-Rashba effect~\cite{bychkov}
and not due to a bulk spin current. This local spin splitting
could result in an electrically-induced spin polarization, similar
to the spatially-homogenous polarization that has been measured in
strained semiconductors~\cite{katoCISP} and semiconductor
heterostructures~\cite{sih,silov}.

In order to clarify the origin of the electrically-induced spin
polarization, we design structures in which the effects of the
boundary of the electric field are separated from edge effects. We
fabricate mesas with transverse channels to allow spins to drift
into regions in which there is minimal electric current. Using
Kerr rotation microscopy, we observe the generation of a
transverse bulk electron spin current created by a longitudinal
voltage, which can cause spins to drift nearly 40 microns into a
transverse channel. In addition, we determine the spin drift
velocity from the magnetic field-dependence of the measured spin
polarization.

The thin-film samples were fabricated from a 2 $\mu$m-thick
epilayer of n-doped GaAs, with a Si doping density $n$ = 3
$\times$ 10$^{16}$ cm$^{-3}$.  The n-GaAs epilayer and the
underlying 2 $\mu$m of undoped Al$_{0.4}$Ga$_{0.6}$As were grown
on a (001) semi-insulating GaAs substrate using molecular beam
epitaxy. Mesas were patterned using photolithography and a
chemical etch, and ohmic contacts were made with annealed AuGe
[Fig. 1(a)].  Each mesa consists of a main channel, fabricated
along the [1$\overline{1}$0] direction along which the electric
field is applied, and two smaller channels that extend from the
main channel in the transverse direction.  The main channels have
length $l$ = 316 $\mu$m and width $w$ = 60 $\mu$m, and the
transverse channels are 40 $\mu$m wide. One mesa has transverse
channels that extend out 10 and 20 $\mu$m from the side of the
channel, and the other mesa has side arms that are 30 and 40
$\mu$m long.

The samples are measured in a low-temperature scanning Kerr
microscope~\cite{stephens} and mounted such that the main channels
are perpendicular to the externally applied in-plane magnetic
field. In order to measure the spin polarization, a linearly
polarized beam is tuned to the absorption edge of the sample
(wavelength $\lambda$ = 825 nm) and is incident upon the sample
through an objective lens, which provides $\sim$1 $\mu$m lateral
spatial resolution. The rotation of the polarization axis of the
reflected beam is proportional to the electron spin polarization
along the beam ($z$) direction.  A square wave voltage with
amplitude ±$V$/2 and frequency 1168 Hz is applied to the device
for lock-in detection. We perform our measurements at a
temperature T = 30 K, and we take the center of the channel to be
the origin.

Kerr rotation is measured as a function of magnetic field (B) and
position ($x$, $y$).  In Figure 1(b), we show data for a position
near the edge of the channel and away from either side arm ($x$ =
26 $\mu$m, $y$ = 0 $\mu$m).  The data is the electrically-induced
spin polarization and can be explained as spin polarization along
the $z$ direction, which dephases and precesses about the applied
magnetic field. This behavior is known as the Hanle
effect~\cite{opticalorientation}, and the curve can be fit to a
Lorentzian function A/[($\omega_{L}$ $\tau_{s}$)$^{2}$ + 1] to
determine the amplitude A and transverse spin coherence time
$\tau_{s}$. The Larmor precession frequency $\omega_{L}$ = g
$\mu_{B}$ B/$\hbar$, where g is the electron g-factor (g = -0.44
for this sample as measured using time-resolved Kerr
rotation~\cite{crooker97}), $\mu_{B}$ is the Bohr magneton, and
$\hbar$ is Planck's constant divided by 2$\pi$.

\begin{figure}
\includegraphics[width=0.44\textwidth]{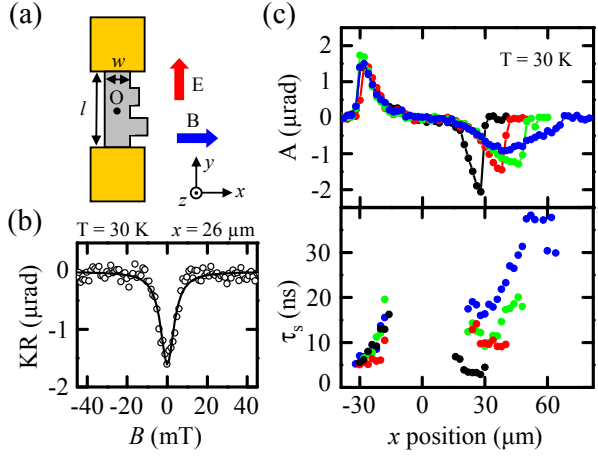}
\caption{\label{fig:epsart} (color) (a) Measurement schematic and
experimental geometry. We take the center of the channel to be the
origin O. (b) Kerr rotation as a function of magnetic field at
($x$,$y$) = (26 $\mu$m, 0 $\mu$m). Line is a Lorentzian fit from
which the amplitude and spin lifetime can be determined. (c) (top)
Spin polarization amplitude as a function of position measured for
the channel (black) and with 10 $\mu$m (red), 20 $\mu$m (green)
and 40 $\mu$m (blue) side arms. (bottom) Spin coherence time as a
function of position. Data are taken at T = 30 K. }
\end{figure}

We repeat this measurement for positions across the channel and
the side arms.  Figure 1(c) shows the amplitude and spin coherence
times for the 60 $\mu$m wide channel and for the channel with 10
$\mu$m, 20 $\mu$m and 40 $\mu$m side arms.  We observe that the
spin polarization amplitude near the edge at $x$ = -30 $\mu$m is
unchanged by the addition of the side arms.  In contrast, the spin
accumulation near $x$ = 30 $\mu$m is modified in the presence of
the side arms, and the spin accumulation has a different spatial
profile that is dependent on the side arm length.  First, we
notice that the spin accumulation is not always largest near the
mesa edge.  This is a clear indication that the spin polarization
is not a local effect caused by the sample boundary.  In addition,
at any position $x$, the amplitude of the spin polarization is
smaller for longer side arms.  This is an indication that we are
observing the spins drifting from the main channel and towards the
end of the side arms.

The magnetic field dependence of the spin polarization is also
different in the side arms.  Using the Hanle model, the width of
the field scans yields the spin coherence time.  Near the edge at
$x$ = -30 $\mu$m, we observe that the spin coherence time appears
to increase for positions closer to the center of the channel, as
reported in Ref.~[\onlinecite{katoSHE}].  In the side arms, the
spin coherence time appears to increase even more for positions
farther from the edge of the channel.  As noted in
Ref.~[\onlinecite{katoSHE}], this spatial dependence could be an
effect on the lineshape due to the time that it takes for the
spins to drift or an actual change in the spin coherence time for
the spins that have diffused from the edge.  Here we will show
that the change in the lineshape in the transverse channels is not
due to an actual change in spin lifetime but can be explained
using a model that incorporates spin drift~\cite{crooker05,lou}.

\begin{figure}
\includegraphics[width=0.44\textwidth]{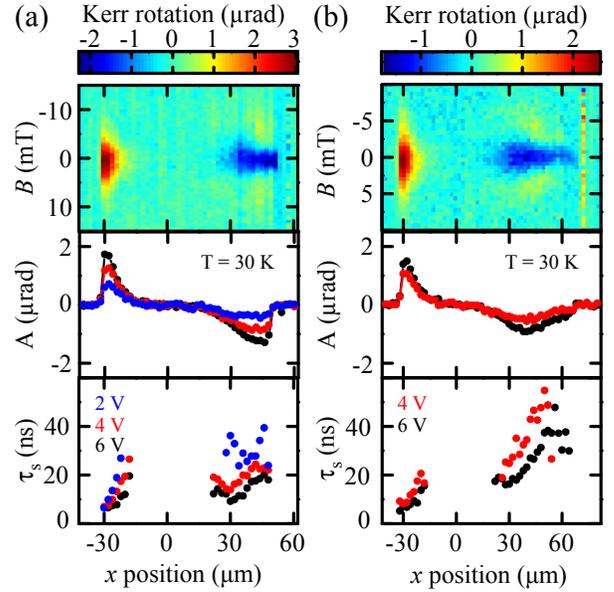}
\caption{\label{fig:epsart} (color) (a) Measurements of the
channel with a 20 $\mu$m side arm at T = 30 K. The edges of the
main mesa are at $x$ = -30 $\mu$m and $x$ = +30 $\mu$m. Top panel
is a color plot of Kerr rotation as a function of magnetic field
and position for $V$ = 6 V. Middle panel shows the amplitude as a
function of position for $V$ = 2 V (blue), 4 V (red) and 6 V
(black). Bottom panel shows the spin coherence time as a function
of position for $V$ = 2 V (blue), 4 V (red) and 6 V (black). (b)
Measurements of the channel with a 40 $\mu$m side arm at T = 30 K.
Top panel is a color plot of spin polarization as a function of
magnetic field and position for $V$ = 6 V. Middle panel shows the
amplitude as a function of position for $V$ = 4 V (red) and 6 V
(black). Bottom panel shows the spin coherence time as a function
of position for $V$ = 4 V (red) and 6 V (black). For the color
plots, a vertical offset is subtracted from each magnetic field
scan. }
\end{figure}

We examine the electric field dependence of the spin accumulation
by performing spatial scans for different voltages.  In Figure
2(a), we present measurements of the channel with a 20 $\mu$m side
arm, and in Figure 2(b), we present measurements of the channel
with a 40 $\mu$m side arm.  The top panel shows the spin
polarization as a function of applied magnetic field and position
for $V$ = 6 V, and the amplitude and spin coherence time are shown
below.  From the color plots, the spatial dependence of the width
of the field scans is apparent.  We find that the amplitude of the
measured polarization increases and the spin coherence time
decreases with increasing voltage.

\begin{figure}
\includegraphics[width=0.44\textwidth]{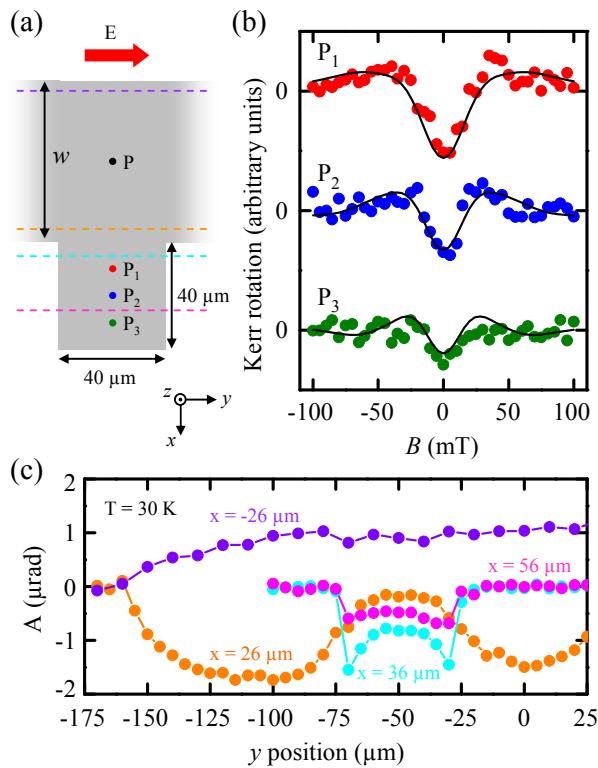}
\caption{\label{fig:epsart} (color) (a) Measurement schematic
showing sample dimensions. P indicates the position ($x$,$y$) = (0
$\mu$m, -50 $\mu$m). (b) Kerr rotation as a function of magnetic
field for three different $x$ positions at $y$ = -50 $\mu$m on the
40 $\mu$m side arm. Taking $x$ = 0 $\mu$m to be the center of the
channel, the edges of the main channel are at $x$ = -30 $\mu$m and
x = +30 $\mu$m ($w$ = 60 $\mu$m). Measurements are shown for $x$ =
40 $\mu$m (P$_{1}$, red), $x$ = 50 $\mu$m (P$_{2}$, blue) and $x$
= 60 $\mu$m (P$_{3}$, green). Solid black lines are calculated
from a model that accounts for spin drift, as described in the
text. The same values for S$_{0}$, $\tau_{s}$, v$_{sd}$ and $D$
are used for all three curves. (c) Amplitude of Kerr rotation
measured as a function of longitudinal position $y$ for $x$ = -26
$\mu$m (purple), $x$ = 26 $\mu$m (orange), $x$ = 36 $\mu$m (cyan)
and $x$ = 56 $\mu$m (magenta). These positions are shown as dashed
lines in part (a) of this figure. One contact edge is located at
$y$ = -158 $\mu$m, and the edges of the side arm are at $y$ = -70
$\mu$m and $y$ = -30 $\mu$m. }
\end{figure}

The Hanle model assumes a constant rate of spin generation, but
this does not accurately reflect what occurs in the side arms,
where there should be minimal electric current.  Instead, we must
consider a model that takes into account the fact that the spins
are generated in the main channel and then drift into the side
arm~\cite{crooker05,lou}.  This signal is computed by averaging
the spin orientations of the precessing electrons over the
Gaussian distribution of their arrival times.  For spins injected
with an initial spin polarization S$_{0}$ along the $z$ direction
at x$_{1}$ and then flow with a spin drift velocity
v$_{\mathrm{sd}}$ before they are measured at a position x$_{2}$,
\begin{align}
S_{z}(x_{1},x_{2},B) = \int_{0}^{\infty} \frac{S_{0}}{\sqrt{4 \pi
D t}} e^{-(x_{2}-x_{1}-v_{\mathrm{sd}}t)^{2}/4 D t}
e^{-t/\tau_{s}} \notag\\ \times \cos(\omega_{L} t) dt
\end{align}
where $D$ is the spin diffusion constant. S$_{z}$($x_{2}$, B) is
computed by integrating $x_{1}$ over the width of the main
channel, from -30 $\mu$m to +30 $\mu$m.  We apply this model to
measurements taken on the 40 $\mu$m side arm, a schematic of which
is shown in Fig. 3(a). Using resonant spin
amplification~\cite{kikkawa}, we determine $\tau_{s}$ = 11.4 ns.
The same set of parameters is used to calculate S$_{z}$($x$, B)
for three positions in the side arm.  As shown in Fig. 3(b), the
model can reproduce the spatial dependence of the amplitude and
lineshape using $D$ = 10 cm$^{2}$/s and v$_{\mathrm{sd}}$ = 1.6
$\times$ 10$^{5}$ cm/s and without assuming a spatially-dependent
spin lifetime. The value obtained for v$_{\mathrm{sd}}$ may have a
contribution from electric field gradients in the transverse
channel, as well as the spin Hall effect.

To check the uniformity of the spin polarization along the
longitudinal ($y$) direction, we perform spatial scans along the
channel and across the side arms.  We show the amplitude as
determined from Lorentzian fits in Figure 3(c).  The contact is
located at $y$ = -158 $\mu$m, and from the scans taken at $x$ =
-26 $\mu$m and $x$ = 26 $\mu$m, we see that the amplitude of the
spin polarization builds up from zero at the contact to a maximum
value over ~50 $\mu$m. While the amplitude of the scans taken at
$x$ = -26 $\mu$m are insensitive to the position of the side arm,
which has edges at $y$ = -70 $\mu$m and $y$ = -30 $\mu$m, the
measurements at $x$ = 26 $\mu$m show that the amplitude drops near
the side arm due to spin drift into the side arm.  Measurements
taken across the side arm, at $x$ = 36 $\mu$m and $x$ = 56 $\mu$m,
show that the amplitude is largest near the edges of the side arm,
which may also be due to spin drift.

\begin{figure}
\includegraphics[width=0.44\textwidth]{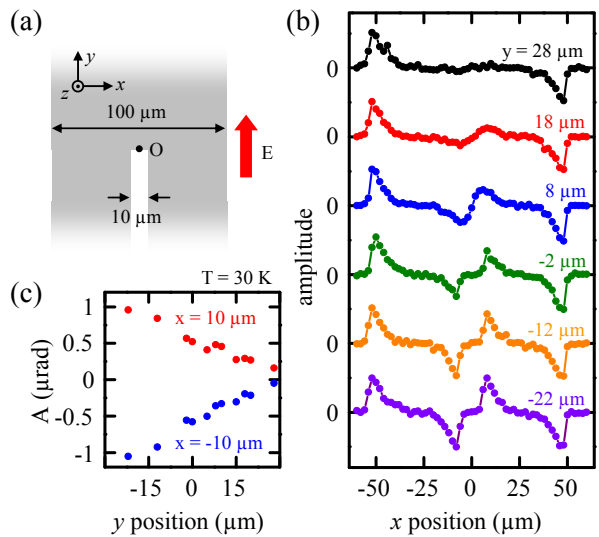}
\caption{\label{fig:epsart} (color) (a) Measurement schematic
showing a channel that splits into two smaller channels. O
indicates the origin. (b) Kerr rotation amplitude as a function of
transverse position $x$ for longitudinal positions $y$ = 28 $\mu$m
(black), $y$ = 18 $\mu$m (red), $y$ = 8 $\mu$m (blue), $y$ = -2
$\mu$m (green), $y$ = -12 $\mu$m (orange) and $y$ = -22 $\mu$m
(purple). The sample edges are at $x$ = -50 $\mu$m, $x$ = -5
$\mu$m, $x$ = 5 $\mu$m, and $x$ = 50 $\mu$m. (c) Kerr rotation
amplitude as a function of $y$ for $x$ = -10 $\mu$m (blue) and $x$
= 10 $\mu$m (red) showing spin diffusion along the direction of
the applied electric field. }
\end{figure}

Finally, we consider a sample geometry that allows us to study
spin drift along the direction of the electric current.  We
fabricate a device in which a 100 $\mu$m channel is split into two
45 $\mu$m arms with a 10 $\mu$m gap [Fig. 4(a)].  In this case, we
take the origin to be the center of the channel in the $x$
direction and where the channel splits into two smaller channels
in the $y$ direction.  We perform measurements of the spin
polarization across the channel, and the data is shown in Fig.
4(b). The maximum spin polarization measured along the inner edges
has the same magnitude as the maximum spin polarization measured
at the outer edges for $y$ = -22 $\mu$m.  This is because the spin
Hall amplitude depends on the electric field but is independent of
the channel width in this regime.  In Fig. 4(c), we plot the
longitudinal dependence of the spin polarization amplitude for $x$
= 10 $\mu$m and $x$ = -10 $\mu$m and observe that the spin
polarization decreases yet persists for nearly 30 $\mu$m into
where the mesa is a single branch. Since these measurements are
performed by locking-in to an oscillating electric field, it is
unknown whether this spin polarization is drifting with or
diffusing against the electric field, but the former case is more
likely.

These measurements demonstrate that the spin Hall effect can drive
transport of spins over length scales that are many times the spin
diffusion length L$_{s}$ = 9 $\mu$m (from fits of the spatial spin
Hall profile to the model in Ref.~[\onlinecite{zhang}]) and with a
transverse spin drift velocity v$_{\mathrm{sd}}$ = 1.6 $\times$
10$^{5}$ cm/s that is comparable to the longitudinal charge drift
velocity v$_{cd}$ = 4.8 $\times$ 10$^{5}$ cm/s at $V$ = 6 V.

We thank Y. K. Kato for discussions and acknowledge support from
ARO, DARPA/DMEA, NSF and ONR.
\\


\begin{thebibliography}{99}

\bibitem{dyakonov} M. I. D'yakonov and V. I. Perel, JETP Lett.
{\bf 13}, 467 (1971)

\bibitem{hirsch} J. E. Hirsch, Phys. Rev. Lett. {\bf 83}, 1834
(1999).

\bibitem{murakami} S. Murakami, N. Nagaosa and S. C. Zhang,
Science {\bf 301}, 1348 (2003).

\bibitem{sinova} J. Sinova, D. Culcer, Q. Niu, N. A. Sinitsyn, T.
Jungwirth and A. H. MacDonald, Phys. Rev. Lett {\bf 92}, 126603
(2004).

\bibitem{zhang} S. Zhang, Phys. Rev. Lett. {\bf 85}, 393 (2000).

\bibitem{katoSHE} Y. K. Kato, R. C. Myers, A. C. Gossard and D. D.
Awschalom, Science {\bf 306}, 1910 (2004).

\bibitem{wunderlich} J. Wunderlich, B. Kaestner, J. Sinova and T.
Jungwirth, Phys. Rev. Lett. {\bf 94}, 047204 (2005).

\bibitem{sih} V. Sih, R. C. Myers, Y. K. Kato, W. H. Lau, A. C.
Gossard and D. D. Awschalom, Nature Physics {\bf 1}, 31 (2005).

\bibitem{engel} H.-A. Engel, B. I. Halperin and E. I. Rashba,
Phys. Rev. Lett. {\bf 95}, 166605 (2005).

\bibitem{tse06} W.-K. Tse and S. Das Sarma, Phys. Rev. Lett. {\bf
96}, 056601 (2006).

\bibitem{nomura} K. Nomura, J. Wunderlich, J. Sinova, B. Kaestner,
A. H. MacDonald and T. Jungwirth, Phys. Rev. B {\bf 72}, 245330
(2005).

\bibitem{rashba} E. I. Rashba, Phys. Rev. B {\bf 70}, 161201(R)
(2004).

\bibitem{shi} J. Shi, P. Zhang, D. Xiao and Q. Niu, Phys. Rev.
Lett. {\bf 96}, 076604 (2006).

\bibitem{tse05} W.-K. Tse, J. Fabian, I. Zutic and S. Das Sarma,
Phys. Rev. B {\bf 72}, 241303(R) (2005).

\bibitem{galitski} V. M. Galitski, A. A. Burkov and S. Das Sarma,
cond-mat/0601677 (2006).

\bibitem{bychkov} Y. A. Bychkov and E. I. Rashba, J. Phys. C {\bf
17}, 6039 (1984).

\bibitem{katoCISP} Y. K. Kato, R. C. Myers, A. C. Gossard and D.
D. Awschalom, Phys. Rev. Lett. {\bf 93}, 176601 (2004).

\bibitem{silov} A. Yu. Silov, P. A. Blajnov, J. H. Wolter, R. Hey,
K. H. Ploog and N. S. Averkiev, Appl. Phys. Lett. {\bf 85}, 5929
(2004).

\bibitem{stephens} J. Stephens, R. K. Kawakami, J. Berezovsky, M.
Hanson, D. P. Shepherd, A. C. Gossard and D. D. Awschalom, Phys.
Rev. B {\bf 68}, 041307(R) (2003).

\bibitem{opticalorientation} {\it Optical Orientation}, F. Meier,
B. P. Zakharchenya, Eds. (Elsevier, Amsterdam, 1984).

\bibitem{crooker97} S. A. Crooker, D. D. Awschalom, J. J.
Baumberg, F. Flack and N. Samarth, Phys. Rev. B {\bf 56}, 7574
(1997).

\bibitem{crooker05} S. A. Crooker, M. Furis, X. Lou, C. Adelmann,
D. L. Smith, C. J. Palmstrom and P. A. Crowell, Science {\bf 309},
2191 (2005).

\bibitem{lou} X. Lou, C. Adelmann, M. Furis, S. A. Crooker, C. J.
Palmstrom and P. A. Crowell, Phys. Rev. Lett. {\bf 96}, 176603
(2006).

\bibitem{kikkawa} J. M. Kikkawa and D. D. Awschalom, Phys. Rev.
Lett. {\bf 80}, 4313 (1998).




\end{thebibliography}
\end{document}